\documentclass{esannV2}
\usepackage{graphicx}
\usepackage[latin1]{inputenc}
\usepackage{amssymb,amsmath,array}
\usepackage[table,xcdraw]{xcolor}
\usepackage{tabularx}

\usepackage{changepage}
\usepackage{amsfonts}
\usepackage{subcaption}
\usepackage{epstopdf}

\begin{document}
%style file for ESANN manuscripts
\title{
Towards Assistive Diagnoses in m-Health:\\A Gray-box Neural Model for Cerebral Autoregulation Index} %Single-neuron Clustering
%Contextual Affordances for Action-Effect Prediction in a Robotic-Cleaning Task}

%***********************************************************************
% AUTHORS INFORMATION AREA
%***********************************************************************
\author{Jorge Cuevas$^1$, Claudio Henr\'iquez$^2$, Francisco Cruz$^{2,3}$
%
% Optional short acknowledgment: remove next line if non-needed
%\thanks{This is an optional funding source acknowledgement.}
%
% DO NOT MODIFY THE FOLLOWING '\vspace' ARGUMENT
\vspace{.3cm}\\
%
% Addresses and institutions (remove "1- " in case of a single institution)
$^{1}$ Departamento de Sistemas Inform\'aticos y Computaci\'on,\\Universitat Polit\`ecnica de Val\`encia, Valencia, Spain.\\
$^{2}$ Escuela de Ingenier\'ia, Universidad Central de Chile, Santiago, Chile.\\
$^{3}$ School of Information Technology, Deakin University, Geelong, Australia.\\

%
% Remove the next three lines in case of a single institution
%\vspace{.1cm}\\
%2- School of Second Author - Dept of Second Author \\
%Address of Second Author's school - Country of Second Author's school\\
}
%***********************************************************************
% END OF AUTHORS INFORMATION AREA
%***********************************************************************

\maketitle

\begin{abstract}
The cerebral autoregulation system (CAS), is a mechanism which aims to regulate pressure variations occurring in the cerebral circulatory system. At present, there only exist invasive methods and, in turn, they are not used to prevent cerebrovascular accidents. Nowadays, the emergent concept of m-Health allows to use mobile devices to assist the cerebral autoregulation index (ARI). For this, it is necessary to find novel models which allow to approximate the ARI by using the blood pressure value. This work proposes a gray-box neural model to find a relation between the arterial blood pressure (ABP) and the cerebral blood flow velocity (CBFV) in order to obtain the ARI. Preliminary results show a good performance by using a phenomenological model in comparison to the Aaslid-Tiecks model.
\end{abstract}

\section{Introduction}

The cerebral autoregulation system (CAS) is one of the fundamental biologic mechanisms. In mammals, the CAS allows the human body to work properly. This system supplies blood to the cerebral region, also providing needed nutrients which are metabolized in the brain. 
In general terms, the cerebral autoregulation is affected by physiological and physicochemical variables, e.g. cerebral metabolic rate, posture, or carbon dioxide levels in the arteries; leading to a non-static autoregulation system and to a highly dynamic system able to adapt to sudden changes of blood pressure. Hence, its correct operation it is fundamental to avoid cerebrovascular diseases and keep a healthy brain. 
Currently, methods to measure and diagnose cerebrovascular diseases are invasive. The skull makes difficult to take directly brain measures, therefore, it is not practical for patients to determine the brain condition using this kind of exams \cite{Chacon}. 

Furthermore, cranial trauma and cerebrovascular diseases are the base for some of the most frequent and dangerous neurological disorders currently detected due to the direct impact in the human brain \cite{Saposnik}. These diseases may be caused by serious skull injuries as well as an interruption of the cerebral blood flow, the latter, due to the clot generation or intense haemorrhage in a blood vessel preventing the normal blood circulation and the supply of oxygen to the brain. This malfunction may cause a brain disorder or even death, therefore, it is fundamental for the human brain to be optimally regulated. 

The cerebral autoregulation index (ARI) is a value fluctuating between 0 and 9 \cite{Aaslid}, which indicates whether a person is doing the cerebral autoregulation properly (0 absence of autoregulation, 9 perfect autoregulation). Currently, there are no exams nor models which compute a precise ARI. This index is very hard to obtain by a simple measure, as aforementioned, there are many variables which influence in the CAS. 

Therefore, there is an opportunity, on the one hand, to study novel methods to obtain variables difficult to measure, and methods which are not fully developed in the state-of-the-art literature \cite{Illodo}. On the other hand, we take into account the growing use of mobile devices which could lead to an assistive diagnose by using smartphones.

\section{The Aaslid-Tiecks model}
To better understand how to compute the ARI, we show the Aaslid-Tiecks (A-T) model. The A-T model uses four state equations to represent changes in the blood pressure P(t). The set of equations is able to obtain the cerebral blood flow velocity (CBFV) \cite{Aaslid} which is represented by V as follows:

\begin{equation}
dP(t) = \frac{\substack{P(t)}}
{\substack{1-CrCP}}
\end{equation}

\begin{equation}
X_1(t) =X_1(t-1) +  \frac{\substack{dP(t-1)-X_2(t-1)}}
{\substack{f \times T}}
\end{equation}

\begin{equation}
X_2(t) =X_2(t-1) +  \frac{\substack{X_1(t-1)-2 \times D \times -X_2(t-1)}}
{\substack{f \times T}}
\end{equation}

\begin{equation}
V'(t) = 1 + dP(t-1) - K \times X_2(t)
\end{equation}
where $dP(t)$ normalizes the pressure using a baseline, $CrCP$ is the critical closing pressure, $f$ corresponds to the sampling frequency, $K$ represents a gain parameter in the equation, $T$ is the time constant and $D$ is the damping factor. Furthermore, $X_1(t)$ and $X_2(t)$ are the state variables of a second-order differential system. 

This proposal of A-T shows ten different theoretical responses according to how are combined the parameters $K$, $D$, and $T$, which are associated with a fixed value of ARI as shown in Table \ref{association}. 
For each measured of $P(t)$, the A-T model produces ten curves representing each ARI based on the velocity $V'(t)$. The curves are compared with the real velocity of the subject and are measured using minimal square error or maximal correlation between the real velocity and the estimated velocity by the model. When the real velocity fits one of the ten estimated curves by either error or correlation, an ARI value is assigned.

\begin{table}[]
\centering
\begin{tabular}{|
>{\columncolor[HTML]{ECF4FF}}c |
>{\columncolor[HTML]{ECF4FF}}c |
>{\columncolor[HTML]{ECF4FF}}c |
>{\columncolor[HTML]{ECF4FF}}c |}
\hline
K & D & T & ARI \\ \hline
0.00 & 1.70 & 2.00 & 0 \\ \hline
0.20 & 1.60 & 2.00 & 1 \\ \hline
0.40 & 1.50 & 2.00 & 2 \\ \hline
0.60 & 1.15 & 2.00 & 3 \\ \hline
0.80 & 0.90 & 2.00 & 4 \\ \hline
0.90 & 0.75 & 1.90 & 5 \\ \hline
0.94 & 0.65 & 1.60 & 6 \\ \hline
0.96 & 0.55 & 1.20 & 7 \\ \hline
0.97 & 0.52 & 0.87 & 8 \\ \hline
0.98 & 0.50 & 0.65 & 9 \\ \hline
\end{tabular}
\caption{Association between K, D, T, and ARI.}
\label{association}
\end{table}

\section{The Simpson$'$s model}

Currently, the optimized Simpson$'$s model \cite{Simpson} to compute the ARI establishes that the relation between the CBFV and the arterial blood pressure (ABP) is represented by the equation (5).

\begin{equation}
V(i)={h(0)p(i)+h(1)p(i-1)+h(2)p(i-2)+...+h(6)p(i-6)}
\end{equation}
where $i$ is the sampling index, $V$ is the CBFV, $p$ is the ABP, and $h$ is the is the resulting coefficient to a filter impulse response (FIR). The filter uses as input and output the CBFV and the APB to get a numeric relation. For ABP, it is posed that exists a set of common coefficients to the subjects. Therefore, having only the pressure values is possible to obtain the estimated flow velocity and thus to compute the ARI based on the model A-T using the equations (1-4).

\section{Proposed model}

We hypothesize that an artificial neural network combined with the phenomenological model described in equation (5), to represent a gray-box model, may lead to a better association between the CBFV and the ABP and, therefore, to better results when computing the ARI.

\begin{figure}%[!t]
  \centering
    \includegraphics[width=\linewidth]{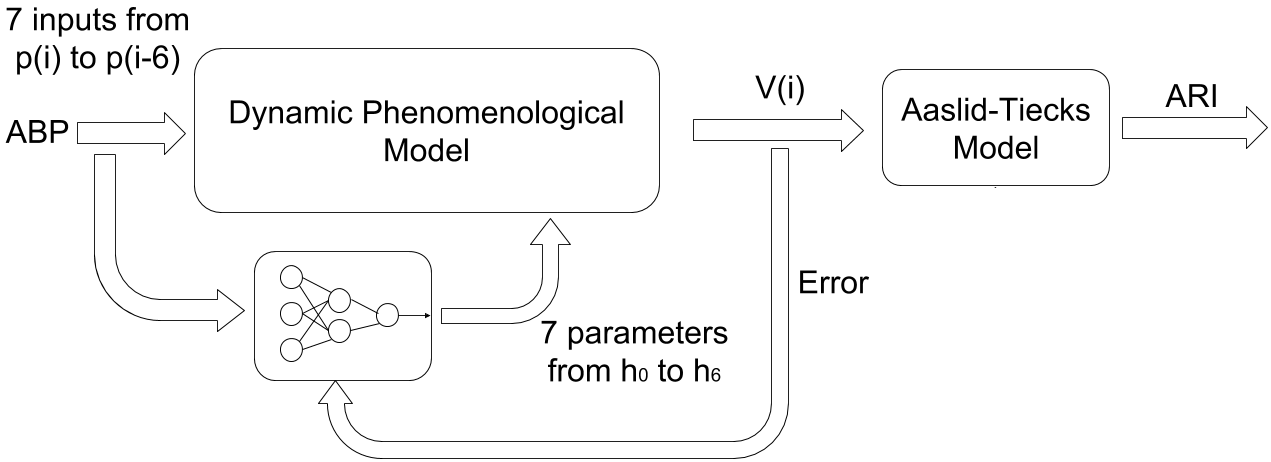}
  \caption{Proposed hybrid model to obtain the ARI.}
  \label{gbneural}
\end{figure}

The proposed model comprises a neural gray-box model within a hybrid model as shown in Fig. \ref{gbneural}. 
The gray-box model is based on an artificial neural network (black box) and a phenomenological part (white box) \cite{Acuna, Cruz2}. 
Furthermore, the hybrid model consists of the gray-box model and the A-T model which uses the variables ABP and CBFV estimated by the neural network to obtain the estimated ARI for a test subject. We propose to replace the coefficients from equation (5) to coefficients estimated by a gray-box neural model.

%Figure. 2

In the model shown in Fig. \ref{gbneural}, it is observed that the hybrid model input is the ABP represented as 7 values which are associated to a flow velocity inside the gray-box model, thereafter it is evaluated in the A-T model to obtain the ARI.
In the hybrid model, the pressures work as a double input, in terms of they are given to the neural network as well as to the proposed A-T model with different aims in each subsystem.
The gray-box model used in this work has been designed to work in a serial manner. This means that the obtained results by the empirical part of the model are sent directly to the phenomenological model \cite{Cruz} as shown in Fig. \ref{gbneural}.

The inputs of the neural network are the ABP of the subjects and, as stated above, they have a direct relationship with a single velocity, therefore, the output is defined as the CBFV.
The training of the gray-box model is performed using indirect training, i.e. the error is computed at the output of the phenomenological model of the gray box \cite{Cruz}. We are interested to know the performance of the whole model and not particularly in the neural part of the model, therefore, we compute the error at the output of the gray-box model using the CBFV.

The phenomenological part of the gray-box neural model contains the equation (5) which is represented along with the neural network using fixed neurons and weights. In this regard, the training algorithm does not modify these connections to keep unaltered its mathematical meaning of the equation.

In parallel, the empirical model is represented by neurons and weights inside the same network. In contrast with the phenomenological part, these weights are adapted over the training.
To complete the hybrid model, the output of the gray-box model is used to compute the ARI based on equations (1), (2), (3) y (4) according to the A-T model.

\section{Experimental setup and results}

\begin{table}[]
\centering
\begin{tabular}{|c|c|c|c|c|c|c|}
\cline{1-3} \cline{5-7}
\cellcolor[HTML]{96FFFB}Subject & \cellcolor[HTML]{96FFFB}Normo & \cellcolor[HTML]{96FFFB}Hiper &  & \cellcolor[HTML]{96FFFB}Subject & \cellcolor[HTML]{96FFFB}Normo & \cellcolor[HTML]{96FFFB}Hiper \\ \cline{1-3} \cline{5-7} 
1                               & 8                             & 8                             &  & 9                               & 9                             & 8                             \\ \cline{1-3} \cline{5-7} 
2                               & 8                             & 8                             &  & 10                              & 8                             & 8                             \\ \cline{1-3} \cline{5-7} 
3                               & 8                             & 6                             &  & 11                              & 7                             & 6                             \\ \cline{1-3} \cline{5-7} 
4                               & 8                             & 8                             &  & 12                              & 7                             & 6                             \\ \cline{1-3} \cline{5-7} 
5                               & 7                             & 7                             &  & 13                              & 7                             & 7                             \\ \cline{1-3} \cline{5-7} 
6                               & 8                             & 8                             &  & 14                              & 6                             & 7                             \\ \cline{1-3} \cline{5-7} 
7                               & 7                             & 6                             &  & 15                              & 8                             & 8                             \\ \cline{1-3} \cline{5-7} 
8                               & 8                             & 7                             &  & 16                              & 8                             & 8                             \\ \cline{1-3} \cline{5-7} 
\end{tabular}
\caption{Variations from normocapnia state to hypercapnia state in 16 subjects.}
\label{hipernormo}
\end{table}

\begin{figure}%[!b]
  \centering
    \includegraphics[width=\linewidth]{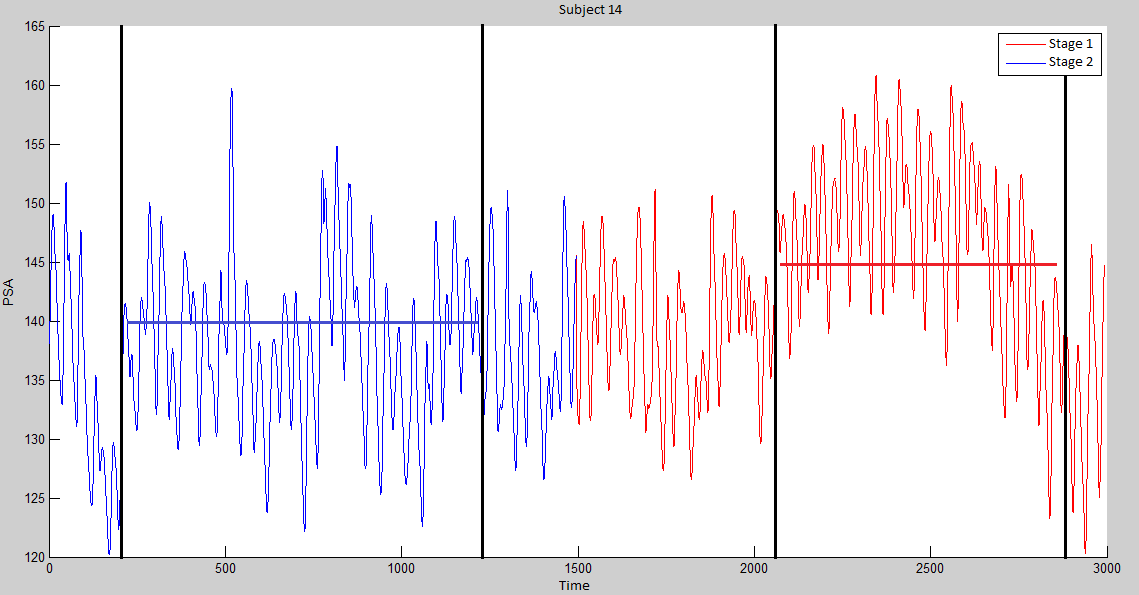}
  \caption{Sample of ABP. Normocapnia in blue (stage 1) and  hipercapnia (stage 2) in red.}
  \label{results}
\end{figure}

The data used in this work were measured in the University of Leicester, England and approved by the ethics committee of the Royal Infirmary Hospital of Leicester. The data comprised the ABP and the CBFV divided in two phases for 16 healthy volunteer patients between 24 and 47 years old. The first phase was carried out with patients in normocapnia state obtaining measures of ABP and CBFV. Afterwards, in the second phase, the subjects were induced to inhale $CO_2$, which led them to a hypercapnia state (as shown in Fig. \ref{results}). The hypercapnia state reduced the oxygen amount in the blood (and thus ARI should decrease), hence, the subjects were not fully healthy as in the first phase. These phases allowed to observe differences through obtained data between the normo- and hypercapnia.
The obtained samples were supervised by a doctor to assure that the data represent actual information for each subject.

The variation between changes of states for each subject should not exceed 2 units, therefore, as shown in Table \ref{hipernormo}, most of the subjects did not present a considerable difference between the ARI values when changing the state. Nevertheless, subject 14 presented an unusual change of the ARI between the states of normocapnia and hypercapnia.

%Figure. 3

%La diferencia entre cambios de estados que debe tener cada sujeto no debe superar 2 valores, por lo tanto, como se senala en la Tabla 2, la mayoría de los sujetos no presenta una amplía diferencia entre los valores del ARI cuando pasan de un estado al otro, sin embargo el sujeto 14 posee un cambio anormal entre normo e hipercapnia.

% ****************************************************************************
% END OF BIBLIOGRAPHY AREA
% ****************************************************************************
\section{Conclusions and Future Work}
The obtained results show that the proposed gray-box neural model is able to correctly identify the coefficients for each subject, showing that there are different p coefficients related to the velocity of the blood flow and blood flow pressure.
The proposed model allows to compute the coefficients for each subject individually, which shows that for a subject A there are 7 different $p$ coefficients to represent the relation between velocity and blood flow pressure.
Moreover, we observed that a few subjects did not change the state from normocapnia to hypercapnia while most of the subjects responded correctly according to the model, excepting for subject 14 who increased his ARI instead of decreased.
%Los resultados obtenidos demuestran que la red neuronal de caja gris es capaz de identificar correctamente los  valores de los coeficientes para cada sujeto, y exhibe que para cada uno de estos existen distintos valores de coeficientes (p) que se relacionan con la velocidad de flujo sanguineo y presion de flujo sanguineo. 
%El modelo permite el calculo individual de coeficientes para cada sujeto, lo que indica que para un sujeto A existen 7 coeficientes p distintos para la representacion de la relacion entre velocidad y presion de flujo sanguineo. 
%Por otra parte, se detectaron sujetos que no cambiaron su estado de normo a hipercapnia, y sujetos que respondieron correctamente al modelo, exceptuando el sujeto 14 que aumento su ARI cuando debio dismunuir. 
As future work, we are planning to develop a mobile app
to assist people to obtain easier diagnoses by mean of an m-Health application.

%The first author gratefully acknowledges the support by \textit{Universidad Central de Chile} and \textit{CONICYT}.

% ****************************************************************************
% BIBLIOGRAPHY AREA
% ****************************************************************************

\begin{footnotesize}

% IF YOU DO NOT USE BIBTEX, USE THE FOLLOWING SAMPLE SCHEME FOR THE REFERENCES
% ----------------------------------------------------------------------------

% ----------------------------------------------------------------------------

% IF YOU USE BIBTEX,
% - DELETE THE TEXT BETWEEN THE TWO ABOVE DASHED LINES
% - UNCOMMENT THE NEXT TWO LINES AND REPLACE 'Name_Of_Your_BibFile'

%\bibliographystyle{unsrt}
%\bibliography{Name_Of_Your_BibFile}

\begin{thebibliography}{00}

\bibitem{Aaslid}
Aaslid, R., Lindegaard, K. F., Sorteberg, W., and Nornes, H. Cerebral autoregulation dynamics in humans. In Stroke, Vol. 20, Nr. 1, pp. 45-52, 1989.

\bibitem{Acuna}
Acu\~na, G., Cruz, F., and Moreno, V., Identifiability of Time varying Parameters in a Grey-Box Neural Model: Application to a Biotechnological Process. In Proceedings of Foodsim2006, Napoles, Italia, 2006.

\bibitem{Chacon}
Chacon, M., Nunez, N., Henriquez, C., and Panerai, R. B. Unconstrained parameter estimation for assessment of dynamic cerebral autoregulation. In Physiological measurement, Vol. 29, Nr. 10, pp. 1179-1193, 2008.

\bibitem{Cruz}
Cruz, F., Acuna, G., Cubillos, F., Moreno, V., and Bassi, D. Indirect training of grey-box models: Application to a bioprocess. In Proceedings of the International Symposium on Neural Networks, pp. 391-397, 2007.

\bibitem{Cruz2}
Cruz, F. and Acuna, G. Indirect training with error backpropagation in gray-box neural model: Application to a chemical process. In Proceedings of the IEEE International Conference of the Chilean Computer Science Society (SCCC), pp. 265-269, 2010.

\bibitem{Illodo}
Illodo Hernandez, D. L., Cruz Toran, D., Gutierrez Gutierrez, D., Martinez Herrer, D., \& Luejes Garcia, D. Estudio de la autorregulacion cerebral en el traumatismo craneoencef\'alico. Utilizacion del test de respuesta hiperemica transitoria. In Revista Cubana de Medicina Intensiva y Emergencias, Vol. 5, Nr. 1, pp. 1-5, 2005.

\bibitem{Saposnik}
Saposnik, G., Del Brutto, O. 2003. Stroke in South America. A systematic review of incidence, prevalence, and stroke subtypes. Storke. 34: 2103-2107 

\bibitem{Simpson}
Simpson, D., Berroeta, C. H., Katsogridakis, E., and Panerai, R. Quantifying autoregulation from estimated model parameters: An optimization approach. In The FASEB Journal, Vol. 28, Nr. 1, pp. 1184-1184, 2014.


\end{thebibliography}

\end{footnotesize}

\end{document}